\documentclass[12pt,preprint]{aastex}

\newcommand{\um}{\ifmmode{\mu{\rm m}}\else{$\mu$m}\fi}
\newcommand{\hh}{\ifmmode{{\rm H}_2}\else{H$_2$}\fi}
\newcommand{\invcm}{\ifmmode{{\rm cm}^{-1}}\else{cm$^{-1}$}\fi}
\newcommand{\twidnu}{\ifmmode{\tilde\nu}\else{$\tilde\nu$}\fi}
\newcommand{\ggt}{GG~Tau}

\newcommand{\hd}{HD~163296}
\newcommand{\aba}{AB~Aur}
\newcommand{\gwo}{GW~Ori}

\newcommand{\kms}{\ifmmode{{\rm km~s}^{-1}}\else{km~s$^{-1}$}\fi}
\newcommand{\escm}{\ifmmode{{\rm ergs~s}^{-1}{\rm cm}^{-2}}\else
                   {ergs~s$^{-1}$~cm$^{-2}$}\fi}
\newcommand{\msol}{\ifmmode{{\rm M}_\odot}\else{M$_\odot$}\fi}
\newcommand{\grapprox}{$_>\atop{^\sim}$}

\shorttitle{Looking for H$_2$ Emission from Disks}
\shortauthors{Richter et al.}
 
\begin{document}

\title{Looking for Pure Rotational H$_2$ Emission \\
from Protoplanetary Disks}

\author{M. J. Richter\altaffilmark{1} and D. T. Jaffe\altaffilmark{1}}
\affil{Astronomy Department, University of Texas, Austin, TX 78712}
\email{richter@astro.as.utexas.edu, dtj@astro.as.utexas.edu}

\author{Geoffrey A. Blake\altaffilmark{1}}
\affil{Divisions of Geological \& Planetary Sciences, \\
California Institute of Technology, MS 150-21, \\
Pasadena, CA 91125}
\email{gab@gps.caltech.edu}

\and

\author{J. H. Lacy\altaffilmark{1}}
\affil{Astronomy Department, University of Texas, Austin, TX 78712}
\email{lacy@astro.as.utexas.edu}

\altaffiltext{1}{Visiting Astronomer at the Infrared Telescope Facility,
which is operated by the University of Hawaii under
contract from the National Aeronautics and Space
Administration.}

\begin{abstract}
We report on a limited search for pure-rotational
molecular hydrogen
emission associated with young, pre-main-sequence stars.
We looked for \hh\ $v=0$ $J=3\rightarrow1$ and $J=4\rightarrow2$
emission in the mid-infrared
using the Texas Echelon-Cross-Echelle
Spectrograph (TEXES) at NASA's 3m Infrared Telescope Facility.
The high spectral and 
spatial resolution of our observations lead to more stringent
limits on narrow line emission close to the source
than previously achieved.
One star, AB Aur, shows a possible (2$\sigma$) \hh\ detection,
but further observations are required to make a confident statement.
Our non-detections suggest that
a significant fraction, perhaps all, of previously reported \hh\ emission
towards these objects could be extended on scales of 5\arcsec\ or more.
\end{abstract}


\section{Introduction}

Formation of a circumstellar disk is recognized as a natural step
in the process of star formation and a vital step toward forming
planets.  
Currently our knowledge of protoplanetary disk properties at anthropically
interesting distances of 1-30~AU is relatively limited.
The best spectroscopic constraints on disk temperature and density
structure are dominated
by material either within 0.5~AU of the central source 
(Najita et al. 1996; Carr, Mathieu, \& Najita 2001)
or at radii $\geq$50~AU 
(Dutrey, Guilloteau, \& Simon 1994).
This radial sampling results
from the tracers utilized to date: dust emission and scattering and CO 
ro-vibrational and rotational emission.  
Using \hh\ rotational lines as a tracer may allow study of disks at
radii of 1-50~AU.  

Recent work by 
Thi et al. (1999, 2001a, 2001b)
report \hh\ $v=0$ $J=2\rightarrow0$ and $J=3\rightarrow1$ emission from 
pre-main-sequence and main-sequence stars with circumstellar disks.
From Infrared Space Observatory (ISO) data, these authors derive
gas temperatures of 100-200~K and 
masses up to 2$\times$10$^{-3}$~\msol.  In the
case of disks around main-sequence stars (Thi et al. 2001a), previously
thought to be mostly gas-free (Zuckerman, Forveille, \& Kastner 1995), the 
presence of a substantial reservoir
of \hh\ alters ideas on the formation of giant planets (Lissauer 2001).
Unfortunately, the large aperture and low spectral resolution of the
observations leave open the question of whether the emission really 
comes from a disk.  

Three mid-IR \hh\ pure rotational lines are readily available from high,
dry sites if observed at moderately high spectral resolution: 
$v=0$~$J=3\rightarrow1$ [{\it i.e.} S(1)] at 17~\um, $J=4\rightarrow2$ 
[{\it i.e.} S(2)]
at 12~\um, and $J=6\rightarrow4$ [{\it i.e.} S(4)] at 8~\um.
The high spectral and spatial
resolution possible with ground-based spectroscopy,
together with large telescope apertures,
can result in greater sensitivity to 
certain gas distributions than satellite observations.  In particular,
ground-based spectroscopy is well suited for detecting point sources
with narrow line emission, such as gas in Keplerian orbit at a radius
of 1-50~AU from a solar-mass star.

We report here on observations of a small sample of young stars
taken with the Texas Echelon Cross Echelle
Spectrograph (TEXES)
in an effort to confirm
the ISO detections and explore the feasibility of ground-based observations
of \hh\ from circumstellar disks.

\section{Observations}\label{sec:obs}

We used TEXES (Lacy et al. 2002) at the 
NASA 3m Infrared Telescope Facility (IRTF) to observe the  
H$_2$~$J=3\rightarrow1$~transition 
[$\lambda=17.0348$~\um\ or $\twidnu=587.032$~\invcm] 
and the $J=4\rightarrow2$ transition 
[$\lambda=12.2786$~\um\ or $\twidnu=814.425$~\invcm].
All the observations were made with TEXES in high-resolution mode.
Pertinent details regarding the observations may be found in 
Table~\ref{tab:obs}.
All sources were observed while using the IRTF offset guide camera as well
as guiding on the dispersed continuum seen through the spectrograph.

The case of \ggt\ deserves special comment since it has weak continuum,
is a strong case for ISO detection 
(Thi et al. 1999; Thi et al. 2001b), and has a 
unique geometry.
The source is a quadruple system composed of a pair of binaries.  
GG~Tau~A is a 0.\arcsec25 binary with millimeter continuum emission 
(Guilloteau, Dutrey, \& Simon 1999)
and HST scattered light observations (Silber et al. 2000)
showing a circumbinary ring 
extending roughly from 180 to 260~AU (1.\arcsec29 - 1.\arcsec86).  
The ring is tipped at
37$^\circ$, resulting in an ellipse on the sky with the major axis 
running essentially E-W.  
When we observed GG~Tau~A, 
we widened the slit to 3\arcsec\ and rotated the instrument to 
orient the slit so that the major axis of the projected ellipse
would lie along our slit.  
The weak continuum of
\ggt\ meant we could not guide on continuum signal
through the spectrograph.  
Therefore, we repeatedly checked our infrared boresight 
by observations of $\alpha$~Tau, a nearby, bright
infrared source.  The maximum boresight offset found based on the
$\alpha$~Tau observations was 0.\arcsec5, with a mean offset of 0.\arcsec2.
By summing over $\approx$4\arcsec\ along
the slit, we can confidently state that we observed essentially all
of the GG~Tau~A gap region.  

Correction for atmospheric transmission was done using bright infrared
continuum objects, either stars or asteroids.  
Although most stars later than spectral type G show photospheric features
at R$\geq20,000$, we can locate features using the Kitt
Peak Sunspot Atlas (Wallace, Livingston, \& Bernath 1994) and the 
ATMOS3 photospheric atlas (Geller 1992) and
know that there are no features near the \hh\ transitions.
Asteroids have no features at our resolution.
Both the $J=3\rightarrow1$ and 
$J=4\rightarrow2$ lines are near telluric atmospheric lines, but the
Doppler shift of the source, the Earth's motion, and the high
spectral resolution available with TEXES helped to minimize
atmospheric effects.  

Flux calibration was done using a blackbody and standard stars.  



\section{Results}\label{sec:res}

The data were reduced using the TEXES data pipeline (Lacy et al. 2002).
We extracted spectra from the data with several strategies: optimal
extraction of the point source to look for emission coincident with
the continuum, sums over the nodded region
to look for diffuse emission covering half the slit length, and selected
sums along the slit to look for isolated emission offset from the 
point source.  We saw no evidence for extended emission in any spectrum,
although uniform emission on \grapprox5\arcsec\ scales would 
not be recovered.

Figure~\ref{fig:s1} presents $J=3\rightarrow1$ spectra for 
the three sources we observed that also
have reported ISO detections: \ggt, \aba, and \hd.  
To indicate the line flux reported by
Thi et al. (2001b), we overplot Gaussians with an integrated line flux equal
to the ISO measurements.  Since ISO was unable to resolve the line profiles,
we have simply assumed Gaussian FWHM matching our resolution (5 or 7.5~\kms,
depending on slit width) 
and 30~\kms, with the
Gaussians centered at the systemic V$_{LSR}$.
The figures clearly show a discrepancy between the reported line flux
and our observations.  

Of all our spectra, only
the \aba\ 12~\um\ spectrum shows a possible detection (Figure~\ref{fig:s2}).
A Gaussian fit to the \aba\ data finds a feature centered at the
systemic velocity with a FWHM=10.5~\kms\
and a line flux of 2.0 ($\pm$1.0) $\times\ 10^{-14}$~\escm.
There are no terrestrial
lines and no photospheric absorption lines in 
the comparison star at this frequency that might artificially
produce an emission feature.
We discuss the implications of
these data, particularly in light of the non-detection of
the $J=3\rightarrow1$ line, below.

Table~\ref{tab:res} summarizes the results for all our observations.
In all cases, we measure continuum levels comparable to past measurements,
although the noise for \ggt\ is such that many pixels must be combined to
obtain a significant continuum measurement.  
We established line flux errors by summing over the number of pixels
corresponding to the assumed line widths
at all parts of the spectrum with comparable atmospheric transmission.
We then use the line flux upper limit for the more conservative case of 
a 30~\kms\ FWHM line to calculate a maximum mass
of \hh\ assuming temperatures of 150~K and 300~K.  Note that these masses
assume no extinction and do not include gas in equilibrium with 
optically thick dust.

\section{Discussion}\label{sec:disc}

We observed six sources
with reasonable evidence for disks and
report no convincing detection of \hh\ pure-rotational emission.
Our non-detections have bearing on the interpretation of previous
ISO reports and on the distribution of gas and dust in these sources.
In all cases where we set limits, those limits are subject to uncertainties
in the assumed gas temperature and the assumed line widths.  
To provide an example of our sensitivity to these assumptions, we include
two line widths and two gas temperatures in Table~\ref{tab:res}.

Three reported ISO detections of \hh\ from circumstellar
disks, out of three tested, are not seen with TEXES.
For ISO to detect emission that we would miss, the emission can either
be extended or it can be spectrally broad.  
Given that the ISO-SWS aperture 
at 17~\um\ covered 14\arcsec\ by 27\arcsec, compared
with our effective aperture of 2\arcsec\ by 2\arcsec\ (280~AU by 280~AU
at Taurus: d=140 pc.), and that
ISO detected \hh\ rotational lines from molecular clouds with even fairly
modest incident UV fields (Li et al. 2002),
the presence
of extended emission is possible.
Thi et al. (2001b) address this issue using $^{12}$CO 3-2 emission and 
conclude that GG~Tau shows no contaminating molecular material, that the
bulk of molecular material for HD~163296 lies in the disk, but that AB~Aur
may have a significant contribution from an extended envelope.
For the case of broad lines, 
we believe we would have seen
lines as broad as $\approx$50~\kms (see Figure~\ref{fig:s1} 
and Table~\ref{tab:res}).
Even for the case of GG~Tau, we would have seen a line as broad as
30~\kms at the 2$\sigma$ level, given the best fit line flux in 
Thi et al. (2001b), although the uncertainties are large enough that these
observations may not be in conflict, especially for a broad line.
Recent \hh\ $v=1\rightarrow0$ $J=3\rightarrow1$ observations
at 2~\um\ show line widths $<$15~\kms\ for GG~Tau
(J. Bary, in preparation),
while CO ro-vibrational emission toward AB~Aur is unresolved at 12~\kms\
(G. Blake, in preparation).  Although the relationship between these
tracers and the pure rotational \hh\ emission depends in detail upon the
excitation mechanism, these observations show that our assumption of
30~\kms\ line widths when calculating mass upper limits is likely conservative.
As a fiducial point, the projected Keplerian
velocity  at a distance of 1 AU around GG~Tau Aa (M = 0.8 M$_{\odot}$; White
et al. 1999) and assuming any circumstellar disk has the same orientation 
as the more extended circumbinary ring (inclination $i=37^{\circ}$), 
would be v=16~\kms.
For AB Aur, taking M=2.4 M$_{\odot}$ (van den Ancker et al. 1998) and 
inclination $i<45^{\circ}$ (Grady et al. 1999), the projected Keplerian 
velocity would be v$<32$~\kms.
Of course, emission lines broadened by Keplerian rotation will generally 
have widths broader than their characteristic Keplerian velocity.
  
Although TEXES is a relatively new instrument,
we feel reasonably confident in
our results.
The successful detection of continuum emission from each source
at levels consistent with past observations means we were
pointed at the source and that our flux determinations are not
dramatically in error.  
We have detected \hh\ rotational emission in sources 
such as
Uranus (L. Trafton, private communication) and NGC 7027 (H. Dinerstein, in 
preparation).
The ISO detections are the result of
special data processing and are near the sensitivity limit 
(Thi et al. 2001b).  Because the ISO reports of molecular gas
reservoirs around main-sequence stars such as $\beta$~Pic
(Thi et al. 2001a) have such important ramifications for our
understanding of planetary formation, the current non-detections
in pre-main-sequence disk sources
strongly argue for follow-up observations of the main-sequence stars
observed by ISO, all of which are reported to be ``medium
confidence'' detections.
Follow-up observations are particularly needed for the well studied
source $\beta$~Pic, where FUSE observations detect no \hh\ absorption
through the edge-on disk
against a stellar O~VI emission line (Lecavelier des Etangs et al.
2001) and where Na~I emission shows a gas disk that seems to coexist 
with the well known dust disk (Olofsson, Liseau, \& Brandeker 2001;
Heap et al. 2000)

In the disk sources we observed, the absence of \hh\ emission certainly
does not mean there is no warm \hh.  Where the dust in the 
disk is optically thick, an
emission line would result only if there is spatial and/or temperature
separation between the dust and \hh, such as a gap in the dust disk
or a hot layer above the optically thick 
disk (D'Alessio et al. 1998; Chiang \&\ Goldreich 1997)
Gaps are believed to be present in \ggt\ and \gwo\ 
(Guilloteau et al. 1999; Mathieu, Adams, \& Latham 1991);
for these sources, the mass upper limits constrain
the gas within the gap.  In general,
our observations argue, albeit weakly, for little spatial separation
or temperature differentiation between the bulk of the \hh\ gas and
dust.

As described in Section~\ref{sec:res}, the 12~\um\ spectrum of AB~Aur 
has a 2$\sigma$ bump at the correct velocity for the $J=4\rightarrow2$ line 
and with a reasonable width.  
When taken with the 3$\sigma$ 
upper-limit on $J=3\rightarrow1$ emission 
from this source, 3.3~$\times\ 10^{-14}$~\escm,
the $J=4\rightarrow2$ line flux of 2.0~$\pm 1.0\ \times\ 10^{-14}$~\escm 
($-1\sigma$, $+1\sigma$)
implies T$_{\rm gas} >$ 380~K (275~K, 500~K), 
assuming no differential extinction.
This is warmer than the 140~K seen
in CO ro-vibrational lines (G. Blake, in preparation). 
For simple thermal emission, we would expect the \hh\ pure rotational lines to
come from cooler regions than CO ro-vibrational emission, although 140~K
gas would not collisionally excite ro-vibrational CO 
emission.  The 3$\sigma$ upper-limit 
on $J=5\rightarrow3$ line flux as observed with 
ISO, 0.4~$\times\ 10^{-14}$~\escm, 
(Thi et al. 2001b) 
combined with the possible $J=4\rightarrow2$ detection 
($-1\sigma$, $+1\sigma$) suggest
T$<$190~K (220~K, 170~K).  
Given the errors and the resulting inconsistency,
we consider the AB~Aur spectrum to give an upper limit to the
$J=4\rightarrow2$ integrated line flux of
4.0~$\times$~10$^{-14}$~\escm.  
As well as reobserving the $J=3\rightarrow1$ and $J=4\rightarrow2$ lines,
ground-based observations
of the $J=6\rightarrow4$ line at 8~\um, 
which should be quite strong if the gas is 
truly at 380~K, would help determine if the $J=4\rightarrow2$ is 
actually as strong
as it appears.

While the next infrared satellites will have exceptional mid-infrared
sensitivity, they may not be able
to detect low line-to-continuum ratio features such as may be 
present in the AB Aur 12~\um\ data.  
If AB Aur 
were observed at R=600, the maximum resolution available with
the SIRTF InfraRed Spectrograph (Roellig et al. 1998), 
a SNR of $\approx$1000 would give a 3$\sigma$ detection of our derived 
line flux.  
The recommended mid-infrared imager/spectrograph for NGST, with
R=1500 (Mather \&\ Stockman 2000),
would require SNR of 450 for a 3$\sigma$ detection.  Higher spectral
resolution on NGST, still a design goal, would dramatically ease this
type of observation.
Future ground-based observations with 8-10m class telescopes and high
resolution spectrographs such as TEXES will improve on the sensitivity 
described here by
an order of magnitude (a factor of 100 in time).

In conclusion, we have attempted to confirm several reports of \hh\
pure-rotational emission from young circumstellar disks without
success.  Our observations generally set more stringent limits on gas within 
$\sim$50~AU of the source and with widths $\leq$50~\kms\ than obtained
with ISO.  We have found possible (2$\sigma$) emission from one source,
AB~Aur, although inconsistencies in derived gas temperatures
lead us to consider the result as an upper limit.
The absence of emission suggests the gas and
dust have little segregation in temperature and/or spatial extent, although
more observations are necessary.  We have shown that
ground-based, high-resolution spectroscopy is 
an effective way to
search for narrow, pure rotational \hh\ line emission such as that expected 
from circumstellar disks.

\acknowledgments

We are happy to acknowledge the day and night staff at the IRTF; without
their hard work, none of this observing would be possible.  We appreciate
the comments of Ewine van Dishoeck and an anonymous referee.  This work was
supported by grant TARP 00365-0473-1999 from the Texas Advanced Research
Program.  MJR acknowledges USRA grant USRA 8500-98-008 through the
SOFIA program.  This research has made use of NASA's Astrophysics Data 
System and the SIMBAD database operated at CDS, Strasbourg, France.


\clearpage

\begin{deluxetable}{cccccc}
\tabletypesize{\scriptsize}
\tablecaption{Observation Information \label{tab:obs}}
\tablewidth{0pt}
\tablehead{
 & 
 & 
\colhead{Resolving} &
\colhead{Slit} & 
\colhead{Slit} &
\colhead{Integ.}  
\\
\colhead{Star} & 
\colhead{Line} & 
\colhead{Power} &
\colhead{Width} & 
\colhead{Length} &
\colhead{Time}  
\\
& 
& 
[R$\equiv \lambda / \delta \lambda $ ] & 
[\arcsec] &
[\arcsec] & 
[sec]  
}
\startdata
GG~Tau & $J=3\rightarrow1$ & 40,000 & 3.0 & 11.5\tablenotemark{a} & 7000 \\
AB~Aur & $J=3\rightarrow1$ & 60,000 & 2.0 & 11.5 & 1800 \\
& $J=4\rightarrow2$ & 83,000 & 1.4 & 7.5 & 2600 \\
HD~163296 & $J=3\rightarrow1$ & 60,000 & 2.0 & 11.5 & 2400 \\
GW Ori & $J=3\rightarrow1$ & 60,000 & 2.0 & 11.5\tablenotemark{a} & 1600 \\
 & $J=4\rightarrow2$ & 83,000 & 1.4 & 7.5 & 3800 \\
L1551 IRS5 & $J=3\rightarrow1$ & 60,000 & 2.0 & 11.5 & 2000 \\
DG Tau & $J=4\rightarrow2$ & 83,000 & 1.4 & 7.5 & 4800 \\
\enddata
 
 
\tablenotetext{a}{The slit was aligned East-West.
Normally, the slit orientation is North-South.}

\end{deluxetable}

\clearpage

\begin{deluxetable}{ccccccccc}
\tabletypesize{\scriptsize}
\tablecaption{Results Summary \label{tab:res}}
\tablewidth{0pt}
\tablehead{

 & 
\multicolumn{2}{c}{ISO Results\tablenotemark{a}} & 
\multicolumn{6}{c}{TEXES Observations}\\

\colhead{Source} & 
\colhead{F$_\nu$} &
\colhead{$J=3\rightarrow1$} & 
\colhead{$\lambda$} & 
\colhead{F$_\nu$\tablenotemark{b}} & 
\multicolumn{2}{c}{Line Flux\tablenotemark{c}} & 
\multicolumn{2}{c}{Mass\tablenotemark{d}} \\

 & 
 & 
\colhead{Line Flux} & 
 & 
& 
5 km/s & 
30 km/s & 
150 K & 
300 K\\

 & 
 & 
\colhead{ [10$^{-14}$} & 
 & 
 & 
 &
 & 
 &
 \\

& \colhead{[Jy]} & 
\colhead{ ergs/s/cm$^{2}$] } & 
\colhead{[$\mu$m]} & 
\colhead{[Jy]} & 
\multicolumn{2}{c}{[10$^{-14}$ ergs/s/cm$^2$]} & 
\multicolumn{2}{c}{[10$^{-5}$ M$_\odot$]} 
}
\startdata
GG~Tau & 1.1 & 2.8 (0.8) & 17 & 0.7 (1.4) & $<$1.6\tablenotemark{e} & $<$3.9 & $<$47 & $<$3.0 \\
AB~Aur & 24.4 & 30 (9) & 17 & 22.7 (2.3) & $<$3.3 & $<$7.2 & $<$91 & $<$5.9 \\
 & -- & -- & 12 & 18.7 (1.3) & $<$4.0\tablenotemark{f} & 
-- & $<$1250\tablenotemark{f} & $<$8.7\tablenotemark{f} \\
HD 163296 & 16.9 & 22 (6) & 17 & 15.3 (2.5) & $<$2.0 & $<$5.1 & $<$46 & 
$<$3.0 \\
L1551 IRS5 & -- & -- & 17 & 20.0 (1.7) & $<$1.4 & $<$4.3 & $<$51 & $<$3.3 \\
GW Ori & -- & -- & 17 & 6.8 (2.0) & $<$1.6 & $<$3.9 & $<$330 & $<$21. \\
& -- & -- & 12 & 6.1 (0.5) & $<$0.5 & $<$2.2 & $<$5300 & $<$37. \\
DG~Tau & -- & -- & 12 & 5.9 (0.5) & $<$0.6 & $<$3.0 & $<$880 & $<$6.2 \\

\enddata
 
\tablenotetext{a}{From Thi et al. 2001b with 1$\sigma$ errors given in 
parenthesis.}

\tablenotetext{b}{Errors given are 1$\sigma$
per pixel.  Continuum level determined
from $>$450 pixels where the atmospheric transmission is reasonably good.}

\tablenotetext{c}{Upper limits are 3$\sigma$ 
assuming Gaussian FWHM of 5 km/s and 30 km/s, except as
noted. }

\tablenotetext{d}{Mass upper limits assume 30~km/s FWHM
Gaussian line profile, except as noted, and assume 
a temperature of 150~K or 300~K.  
}

\tablenotetext{e}{Assuming FWHM=7.5 km/s due to 3\arcsec\ slit.}

\tablenotetext{f}{Derived from fit to feature at systemic velocity
plus 2$\sigma$ (see text).  Fitted FWHM=10.5 km/s.}

\end{deluxetable}


\clearpage


\figcaption[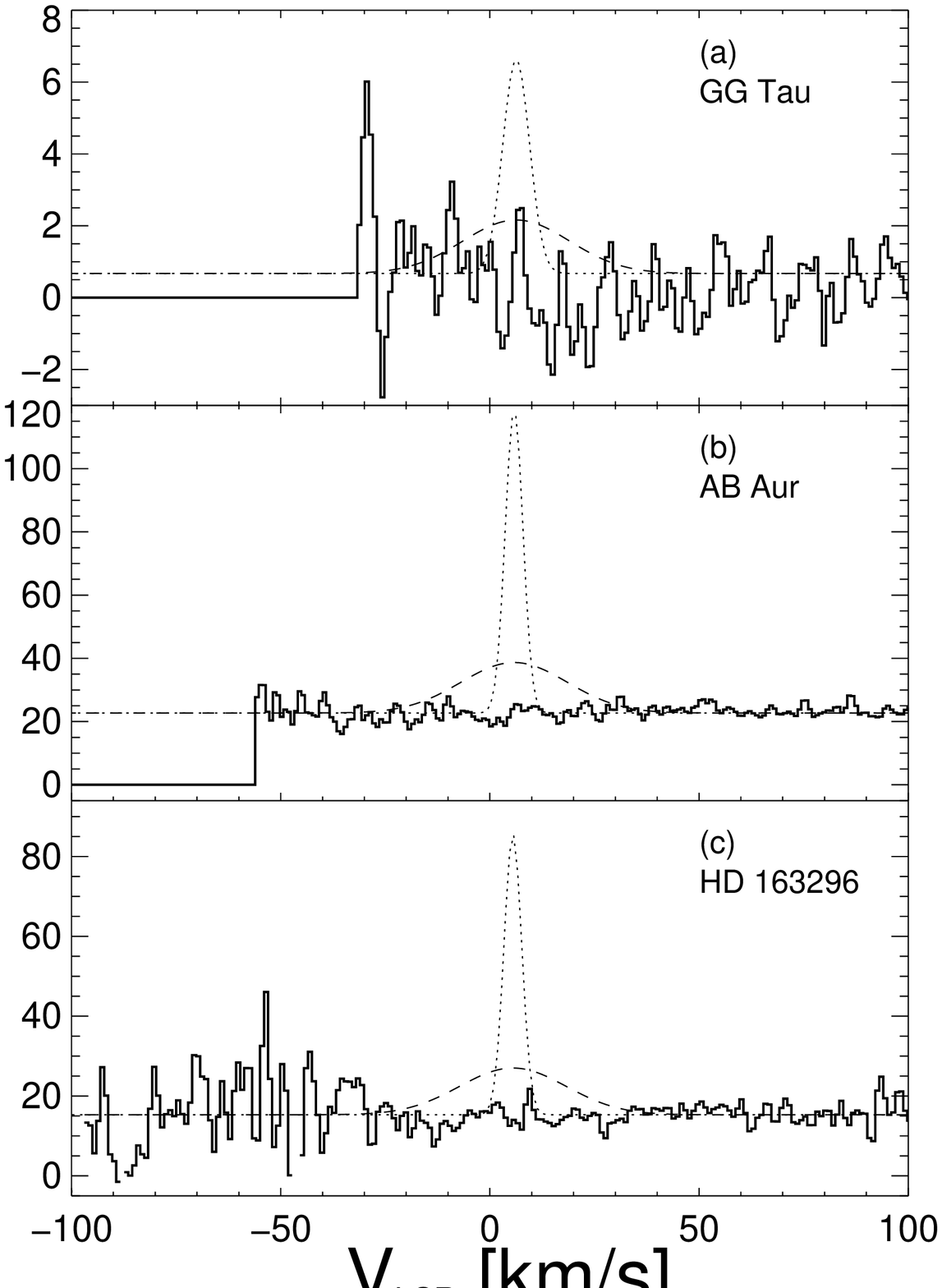]{H$_2$ $J=3\rightarrow1$ ($\lambda=17.035$~$\mu$m)
observations of the three sources observed by TEXES with reported
ISO detections: (a) GG~Tau; (b) AB Aur; and (c) HD 163296.
In all cases, the data (histograms) and overplotted with Gaussians 
that have integrated line flux equal to that reported in Thi et al. (2001b).  
The narrow Gaussian (dotted)
matches our instrumental resolution, FWHM=7.5 km/s for GG~Tau and
FWHM=5 km/s for AB~Aur and HD~163296.  The wide Gaussian (dashed) has
FWHM=30 km/s.  The Gaussians are centered on the systemic velocity.
At this wavelength, the spectral order is larger than the detector;
regions of the spectrum set to 0 Jy are not sampled.
The increased noise toward negative velocities is due to increasing
telluric opacity.
\label{fig:s1}}

\figcaption[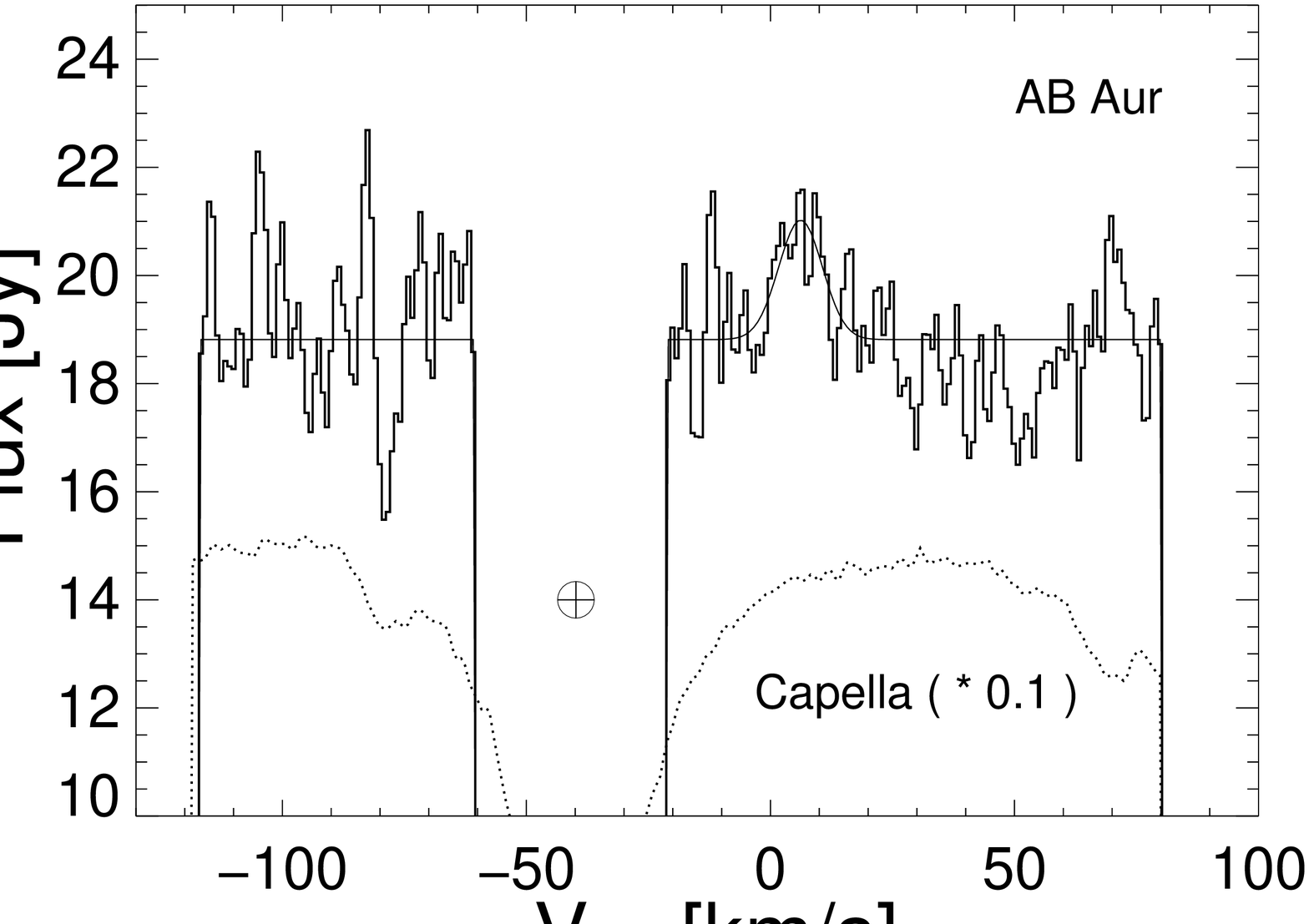]{H$_2$ $J=4\rightarrow2$ 
($\lambda=12.279$~$\mu$m) spectrum
of AB Aur.  One complete order (out of 8) is shown, but the 
region near -40 km/s
where the terrestrial atmospheric transmission is less than 75\%\ 
is set to zero.  The data (histogram) are overplotted with a Gaussian
fit.  The resulting fit has FWHM=10.5~\kms\ and centroid at the
systemic velocity.  The dotted line shows the spectrum of Capella 
(divided by 10), the atmospheric calibrator.  The
features in Capella near -80~km/s and +70~km/s are 
photospheric OH absorption.  Although
we attempted to correct for the OH absorption before
dividing AB~Aur by Capella, some contamination may
still be present.
\label{fig:s2}}

\clearpage

\begin{figure}
\plotone{f1.eps}
\end{figure}

\begin{figure}
\plotone{f2.eps}
\end{figure}


\clearpage

\begin{thebibliography}{}

%
\bibitem[Carr, Mathieu, \& Najita(2001)]{carr01}      
  Carr, J.~S., Mathieu, R.~D., \& Najita, J.~R.\
  2001, \apj, 551, 454

\bibitem[Chiang \& Goldreich(1997)]{chiang97}   
  Chiang, E.~I.~\& Goldreich, P.\ 1997, \apj, 490, 368.

\bibitem[D'Alessio, Canto, Calvet, \& Lizano(1998)]{dalessio98}	
  D'Alessio, P., Canto, J., Calvet, N., \& Lizano, S.\ 1998, \apj, 500, 411. 

\bibitem[Dutrey, Guilloteau, \& Simon(1994)]{dutrey94}  
  Dutrey, A., Guilloteau, S., \& Simon, M.\ 1994, \aap, 286, 149

\bibitem[Geller(1992)]{geller92}        
  Geller, M.\ 1992, Washington, D.C.~: National Aeronautics 
  and Space Administration, Scientific and Technical Information Service, 
  1992., Vol. III

\bibitem[Grady et al.(1999)]{grady99} 	
  Grady, C.~A., Woodgate, B., Bruhweiler, F.~C., Boggess, A., 
  Plait, P., Lindler, D.~J., Clampin, M., \& Kalas, P.\ 1999, \apjl, 523, L151. 

\bibitem[Guilloteau et al. 1999]{guil99}        
  Guilloteau, S., Dutrey, A., \& Simon, M.\ 1999, \aap, 348, 570

\bibitem[Heap et al.(2000)]{Heap00}             
  Heap, S.~R., Lindler, D.~J., Lanz, T.~M., Cornett, R.~H., 
  Hubeny, I., Maran, S.~P., \& Woodgate, B.\ 2000, \apj, 539, 435

\bibitem[Lacy et al.(2002)]{lacy02}     
  Lacy, J.~H., Richter, M.~J., Greathouse, T.~K., Jaffe, 
  D.~T., \& Zhu, Q.\ 2002, \pasp, 114, 153
 
\bibitem[Lecavelier des Etangs et al.(2001)]{lecav01}   
  Lecavelier des Etangs, A.~et al.\ 2001, \nat, 412, 706

\bibitem[Li et al.(2002)]{li02}			
  Li, W., Evans, N.~J.~II, Jaffe, D.~T., van Dishoeck, E.~F., \& 
  Thi, W.~F.\ 2002, \apj, in press

\bibitem[Lissauer(2001)]{lissauer2001}     
  Lissauer, J.~J.\ 2001, \nat, 409, 23
 
%
\bibitem[Mather \& Stockman(2000)]{2000ISASS..14..203M} 	
  Mather, J.~C.~\& Stockman, H.~S.\ 2000, The Institute of Space and 
  Astronautical Science Report SP No.~14, p.~203-209., 14, 203

\bibitem[Mathieu, Adams, \& Latham(1991)]{mathieu91}    
  Mathieu, R.~D., Adams, F.~C., \& Latham, D.~W.\ 
  1991, \aj, 101, 2184

\bibitem[Najita et al.(1996)]{najita96}            
  Najita, J., Carr, J.S., Glassgold, A.E., Shu, F.H., \&
  Tokunaga, A.T. 1996, ApJ, 462, 919

\bibitem[Olofsson, Liseau, \& Brandeker(2001)]{olof01}	
  Olofsson, G.~;., Liseau, R.~;, \& Brandeker, A.\ 2001, \apjl, 563, L77. 


\bibitem[Roellig et al.(1998)]{roellig98} 	
  Roellig, T.~L.~et al.\ 1998, \procspie, 3354, 1192

\bibitem[Silber et al. 2000]{silber00}          
  Silber, J., Gledhill, T., Duch{\^e}ne, G., \& M{\'e}nard, F.
  2000, \apjl, 536, L89

\bibitem[Thi et al.(1999)]{thi99}     
  Thi, W., van Dishoeck, E.~F., Blake, G.~A., van Zadelhoff,
  G., \& Hogerheijde, M.~R.\ 1999, \apjl, 521, L63

\bibitem[Thi et al.(2001a)]{thi01a}      
  Thi, W.~F.~et al.\ 2001a, \nat, 409, 60

\bibitem[Thi et al.(2001b)]{thi01b}         
  Thi, W.~F.~et al.\ 2001b, \apj, 561, 1074

\bibitem[van den Ancker, de Winter, \& Tjin A Djie(1998)]{vdancker98} 
  van den Ancker, M.~E., de Winter, D., \& Tjin A Djie, 
  H.~R.~E.\ 1998, \aap, 330, 145.     

\bibitem[Wallace, Livingston, \& Bernath(1994)]{wallace94}      
  Wallace, L., Livingston, W., \& Bernath, P.\ 1994,
  NSO Technical Report, Tucson: National Solar Observatory, National
  Optical Astronomy Observatory, |c1994,

\bibitem[White et al. 1999]{white99}            
  White, R. J., Ghez, A. M., Reid, I. N., \& Schultz, G. 1999, \apj, 520, 811

\bibitem[Zuckerman, Forveille, \& Kastner(1995)]{zuckerman95}   
  Zuckerman, B., Forveille, T., \& Kastner, J.~H.\ 1995, 
  \nat, 373, 494
 
\end{thebibliography}
\end{document}